\documentclass[12pt]{article}
\usepackage{amsmath,amsfonts,amssymb,latexsym}
\usepackage{epsfig}
\usepackage{graphicx}
\usepackage{wrapfig}
\usepackage{subfigure}
\usepackage{cite}

\def\be{\begin{equation}}
\def\ee{\end{equation}}
\def\bea{\begin{eqnarray}}
\def\eea{\end{eqnarray}}
\def\bean{\begin{eqnarray*}}
\def\eean{\end{eqnarray*}}
\def\C{\mathbb{C}}
\def\I{\mathbb{I}}

\def\R{\mathbb{R}}

\def\D{\mathbb{D}}

\begin{document}

%%%%%%%%%%%%%%%%%%%%%%%%%%%%%%%%%%%

\thispagestyle{empty}
\hfill \today
\vspace{2cm}
\begin{center}
\bf{\LARGE {Theory of Images and Quantum Mechanics,\\
\vspace{.2cm} a common Paradigm\footnote{Contr to 9th Int Conf DICE2018, Sept17-21 2018, Castiglioncello (Italy). See {\it J Phys Conf Ser}}} }
\end{center}
\vspace{1.5cm}

\begin{center}

\bf{\large Enrico  Celeghini}
\end{center}

\begin{center}
{\sl Dipartimento di Fisica, Universit\`a  di Firenze and
INFN--Sezione di
Firenze \\
I50019 Sesto Fiorentino,  Firenze, Italy}\\
\medskip
{\sl Departamento de F\'{\i}sica Te\'orica, Universidad de
Valladolid, \\ 
E-47011, Valladolid, Spain}\\
\medskip

{e-mail: celeghini@fi.infn.it}

\end{center}
\vspace{.6cm}

%%%%%%%%%%%%%%%%%%%%%%%%%%%%%%%%
\begin{abstract} 
The Paradigms introduced in philosophy of science one century ago are shown to be
quite more satisfactory of that introduced by Galileo. This is particularly evident in the physics based on Hilbert Spaces and related mathematical structures that we apply in this paper to Quantum Mechanics and to Theory of Images.
An exhaustive discussion, that include the algebraic analysis of the 
operators acting on them, exhibits that the Hilbert Spaces -that have fixed
dimension- must be generalized to the Rigged Hilbert Spaces that
contains right inside spaces with continuous and discrete dimensions. This is the
property of Rigged Hilbert Spaces that allows a consistent formal description of the
physics we are considering.
Theory of Quantum Mechanics and of Images are similar and the fundamental  difference between 
them come from the definition of measure that is outside the theory of the spaces: while in Quantum Mechanics the measure is a probabilistic action, in Images it is a classical functional.
\end{abstract}
%%%%%%%%%%%%%%%%%%%%%%%%%%%%%%%%%%%%%%%%%%

\bigskip

\centerline{KEYWORDS: Optics, Quantum Mechanics, Rigged Hilbert Spaces, Lie Algebras}
\bigskip

\vfill\eject

%%%%%%%%%%%%%%%%%%%%%%%%%%%%%%%%%%%%%%%%%%

\section{Introduction}

Paradigm in philosophy of science is the {\it Gestalt}, the general frame, that define the
fundamental attitudes of scientists active in  similar fields.

The first general paradigm in science has been stated by Galileo: it could be summarized in the statement
that experience is necessary and sufficient to find the "Truth".
The paradigms of the philosophy of science of the XX century are less arrogant and more problematic. 
The author point of view can
be summed up in the Wittgenstein statement:
{\it "The procedure of induction consists
in accepting as true the simplest law that can be reconciled with our experience"} 
\cite{Witt}.
This means that a law is considered a recipe that worked -up to now- every time has been taken into account and, between the working recipes, is the {\it "simplest"} one.

The role of the general paradigm is relevant in physics, where we 
have at least three conflicted sectorial paradigms that, at the moment, we have not been able to reduce
to a single one. They are General Relativity, Classical Physics and theories related to Hilbert Spaces
with their operators. It is normally assumed that in the appropriate
limit all of them reduces one to an other but, for the time being,
this is more an act of faith that a scientific truth.

Moreover in physics the word {\it "Paradigm"} is almost always substituted  
by the word {\it "Theory"} 
that  is less general but suggests a strong connection with {\it "Truth"} and
the philosophical attitudes of XX century are not too popular since the analysis follows meanly
the mental framework of Galileo:
the physicists  tend to refuse to discuss about the meaning of the theories believing
that physical concepts have in themselves a content of reality.

In this paper we consider only the Hilbert Space paradigm and its applications 
to Quantum Mechanics (QM) and Theory of Images (TI). We discuss as the
Hilbert spaces (HS) -where the dimension is fixed- are insufficient to describe
consistently these applications where operators with spectrum of dimension $\aleph_0$ (discrete) and  $\aleph_1$ (continuous) appear together. We thus introduce a generalization of HS,
the Rigged Hilbert Spaces (RHS), that contains simultaneously spaces of dimensions 
$\aleph_0$ and $\aleph_1$\cite{ReSi, Bo, GaGo, Ce15, CeGaOl}.

RHS -and not HS- are indeed, in the opinion of the author, the more appropriate specific scheme where QM and TI can be formulated. To read this paper the copious 
mathematical literature on RHS (see, for instance, ref.\cite{ReSi}) is not necessary: 
at the level of
applicative aspects RHS are quite similar to HS, except for the fact that inside RHS 
operators and bases of dimension $\aleph_0$ and $\aleph_1$ can be considered together\cite{Ce17}.

Summarizing, this paper is focused on the applications of RHS to QM and TI.
The conclusion is that both applications need similar RHS. A fundamental difference
between QM and TI exists but it is not at the level of the spaces but at the level of something added to the scheme from outside RHS: it is the definition of the measure that is probabilistic in QM and described by deterministic functionals in TI.
\vspace{.3cm} 

%%%%%%%%%%%%%%%%%%%%%%%%%%%%%%%%%%%%%%%%%%
\section{Quantum Mechanics Hilbert Space}
%%%%%%%%%%%%%%%%%%%%%%%%%%%%%%%%%%%%%%%%%%

Hilbert Space is a generalization of inner product vector space. It has fixed
number of dimensions and this is the reason we consider in this paper  the extension to RHS.
The dimensions normally considered in Quantum Mechanics are {$\aleph_0$} (the dimension of
natural numbers) and { $\aleph_1$} (the dimension of the continuum).
 
 Essentials of Quantum Mechanics are:
 
 1)~Any physical state can be exhaustively characterized by a vector.
 
 2)~All observables correspond to Hermitian operators.
 
 3)\hspace{.1cm}The measure process is absolutely different from all others processes.
 
 While the points 1) and 2) are the same in QM and TI, 
 the point 3) is absolutely dissimilar, because QM is a quantum theory and TI a classical one.  Let us thus consider this point 3) of QM in slight more detail.
  
3a) The value of one observable on one of the eigenstates of its corresponding operator is equal to the corresponding eigenvalue.

3b) The value of one observable on a generic state is not fixed but probabilistic and restricted to the eigenvalues of the corresponding operator. The probability of each eigenvalue is proportional to the modulus square of the related component.

As stated by Bohr the measure is something added to the theory of HS. The definition of measure in QM 
is probabilistic and independent from the formal theory of HS because it describes a peculiar action
that implies an interaction of the physical microscopic state with a macroscopic object.
All the rest of QM has nothing to do with probability.

Theory of Quantum Mechanics, as it is, is not actually considered too satisfactory. 
Let us recall the Feynman's authoritative citations:

{\it "I think I can safely say that nobody understands Quantum Mechanics"}\cite{Fey}.

The point is that physicists pose a lot of questions like:

-~ Is the wave function a real object?

-~ Do measurement generates many-worlds?

-~ What is the Copenhagen statistical collapse?

-~ What means the Born rule for repeated measurement?
\vspace{.1cm}

All these questions are perfectly reasonable inside the philosophy of the XVI-XVIII centuries but
a complete {\it nonsense}\, in the spirit of the contemporary theory of science (see, for instance, \cite{Witt, Popp, Bhas}) where  a law, a physical law,  is related to the known experimental results and isn't an absolute. Therefore in this spirit Quantum Mechanics is a magnificent recipe because it describes our electronic apparatuses and it has been -up to now- in agreement with all the experimental results, but it is only a recipe:  there is not any ontological content in its postulates.

To the author opinion, the formal limit of standard QM based on HS is that -to consider a particular case- it is unable to describe together, in a mathematically acceptable way,
energy and position operators of the harmonic oscillator. In general we believe that the essential problem of HS in QM is that they have fixed dimensions, while we 
have to consider together operators with spectrum of different dimensions.
This point will be discussed in next Sections and the proposed solution is to move from HS to the {\it "simpler"}\cite{Witt} RHS.
\vspace{.4cm}

%%%%%%%%%%%%%%%%%%%%%%%%%%%%%%%%%%%%%%%%%%
\section{Rigged Hilbert Space on the line Quantum Mechanics}
%%%%%%%%%%%%%%%%%%%%%%%%%%%%%%%%%%%%%%%%%%

Rigged Hilbert Space on the line $\R$ is defined by\cite{Ce15, CeGaOl}:
\vspace{.1cm}

a)~Two continuous bases $\{|\,x\,\rangle\}$ and $\{|\,p\,\rangle\}$ in\, $\R$ with properties:

\[
\langle\, x\,|\,x'\,\rangle\; = \sqrt{2 \pi}\;\, \delta(x - x')\;,\qquad
\quad \frac{1}{\sqrt{2 \pi}}\int_{-\infty}^{+\infty}\, |x\rangle\, dx\, \langle x|\;  =\; {\mathbb I}\; ,
\]

\[
|p\rangle:=\; \frac{1}{\sqrt{2 \pi}}
\int_{-\infty}^{+\infty}\, dx\; e^{+{\bf i} p x}\,|x\rangle\
\qquad \Rightarrow\qquad 
\langle\, p\,|\,x\,\rangle\; = \; e^{-{\bf i} p x}\,,
\]

\[
\langle\, p\,|\,p'\,\rangle\; = \sqrt{2 \pi}\; \delta(p-p')\,,\qquad\qquad
\quad \frac{1}{\sqrt{2 \pi}}\;\int_{-\infty}^{+\infty} |p\rangle\, dp\, \langle p|\, =\, {\mathbb I}\,.
\]
\vspace{.2cm}

b)~Two bases in $L^2(\R)$ -the space of square integrable functions defined in $\R$- the Hermite functions $\{\psi_n(x)\}$ and $\{\psi_n(p)\}$\,:

\[
\psi_n(x):=\; \frac {e^{-x^2/2}}{\sqrt{2^n n! \sqrt{\pi}}}\; H_n(x)\,,
\]

\[
\psi_n(p):=\;
(-{\bf i})^n [FT]\, \psi_n(x)\;=\; \frac{(-{\bf i})^n}{\sqrt{2 \pi}}
\int_{-\infty}^{\infty}\, dx\; e^{\,{\bf i} p x}\; \psi_n(x) 
\]
(where $H_n$\, are the Hermite polynomials). The   $\{\psi_n\}$   have indeed the
following properties:

\[
\int_{-\infty}^{\infty} \psi_n(x)\; \psi_{n'}(x)\; dx = \delta_{n,n'}\;,\qquad\quad
\sum_{n=0}^\infty\,  \psi_n(x)\;\, \psi_n(x')\, =\; \delta(x-x')\,,
\]

\[
\int_{-\infty}^{\infty} \psi_n(p)\; \psi_{n'}(p)\; dp = \delta_{n,n'}\;,\qquad\quad
\sum_{n=0}^\infty\,  \psi_n(p)\;\, \psi_n(p')\, =\; \delta(p-p')\,.
\]
Note that the Kronecker delta and the Dirac delta appear
 together showing that both the dimensions $\aleph_0$ and $\aleph_1$ are present in $L^2(\R)$ 
that thus cannot be a HS.
\vspace{.1cm}   

c) ~One discrete basis in the line   $\{|n \rangle \}$:
\vspace{.2cm}

Let us define the vector $|n\rangle\in\R$
\[
{ |n\rangle\; :=\; (2 \pi)^{-1/4}\,\int_{-\infty}^{\infty}\, dx\;\, \psi_n(x)\; |x\rangle}\; .
\]
By inspection, the discrete set\, $\{|n \rangle \}$\, is orthonormal and complete
\[
\langle\, n\,|\,n'\,\rangle
= \delta_{n\,n'}\;\;,\qquad\qquad 
\sum_{n=0}^{\infty}\; |n\rangle\, \langle n|\;  =\; { \mathbb I'}\,.
\]
 $\{ |n\rangle \}$ is thus another basis on the real line $\R$,
but with { cardinality $\aleph_0$}. As

\[
 \langle n | x \rangle \;=\; \langle x | n \rangle\; =\;\, (2 \pi)^{1/4}\;  \psi_n(x)\,,
\]
\[
\langle n | p \rangle \; =\; {\bf i}^n\; (2 \pi)^{1/4}\; \psi_n(p)\,, \qquad\quad
\langle p | n \rangle\; =\; (-{\bf i})^n\; (2 \pi)^{1/4}\; \psi_n(p)\,, 
\]
the Hermite functions are nothing else that the transformation matrices between
the continuous bases and the discrete one in\, $\R$.
\vspace{.4cm}

d) ~The Weyl-Heisenberg algebra (WHA):
\vspace{.1cm}

Let us introduce in the space $\{\psi(x)\}$ the operators X, P, N (that read position, momentum and order of the Hermite functions) and 
the identity\, {\bf I}. 
Taking into account the recurrence relation of the Hermite polynomials, we can write
\[
 ~~N\;\, \psi_n(x) \,=\, n\;\, \psi_n(x)\,,\qquad {\bf I}\;\, \psi_n(x) \,=\, \psi_n(x)\,,
 \] 
 \[
  a\;\, \psi_n(x) \;=\; \sqrt{n}\;\, \psi_{n-1}(x) \,,\qquad a^\dagger\;\, \psi_n(x) =
 \sqrt{n+1}\;\, \psi_{n+1}(x)\;;
 \] 
 where
 \[
 a = \frac{1}{\sqrt{2}} (X + {\bf i} P) \;,\;\;\;\;\;\; a^\dagger = \frac{1}{\sqrt{2}}
 (X - {\bf i} P) .
 \]

By inspection $N, a, a^\dagger$ and ${\bf I}$ \,close the Weyl-Heisenberg algebra
\[
[ N, a ] = -a\;,\;\;\;\;\;\; [ N, a^\dagger ] = ~a^\dagger \;,\;\;\;\;\;\; [a, a^\dagger]=  {\bf  I} \;,\;\;\;\;\;\; 
[{\bf I},\bullet ]=0\;;
\]
with Casimir invariant $\C = (N+1/2) {\bf I}-\frac{1}{2}\{a,a^\dagger\}$.
$\{\psi_n(x)\}$ is isomorphic to $\{|n\rangle\}$\, and define the
representation of WHA with Casimir $\C=0$
that is irreducible: the isomorphism between
the Universal Enveloping Algebra of the WHA, $UEA[h(1)]$, and the 
space of the linear operators acting on  $\{|n\rangle\}$ can be extended to $\{\psi_n\}$ 
i.e. every linear operators on $\R$ and $L^2(\R)$ can be written (with $\alpha,\beta,\gamma$
integers and $c_{\alpha,\beta,\gamma}$ complex numbers) as
\[
{\cal O} \;=\; \sum_{\alpha,\beta,\gamma} c_{\alpha, \beta, \gamma}\;\,
(a^\dagger)^\alpha \,N^\beta \,a^\gamma\;\,\,.
\]
The representation is also unitary: every transformation of\, $\{\psi_n\}$\, under the
group is a basis. 

%%%%%%%%%%%%%%%%%%%%%%%%%%%%%%%%%%%%%%%%%%
\section{Rigged Hilbert Space on the disk of Images}
%%%%%%%%%%%%%%%%%%%%%%%%%%%%%%%%%%%%%%%%%%

Theory of Images is quite similar to the theory of Quantum
Mechanics exposed above. Again we have a RHS, constructed now on the 
disk\, $\D$. As told before,
the fundamental difference is not at the level of Rigged Hilbert Space, but
at the level of measure that is defined in a completely different way: 
as Theory of Images is a classical theory, the measure doesn't operate on 
images but is realized by means of functionals. 
\vspace{.3cm}

Rigged Hilbert Space of Images is defined on the disk $\D$ by\cite{Ce17, CeGaOl2}:
\vspace{.1cm}

~a)  ~A continuous basis in the disk\,  $\{|r,\phi \rangle\}$  

A point in the disk is characterized by two coordinates. By means of
the two continuous variables $\{r, \phi\}$ we define a continuous basis
$\{|r,\phi\rangle\}$ (that corresponds to the $\{|x \rangle \}$ of $\R$):
\[
\langle r_1,\phi_1|r_2,\phi_2\rangle \,=\; 2\pi
\; \delta({r_1}^2-{r_2}^2)\;  \delta(\phi_1 - \phi_2)\,,\;\;\;\;\;\;\;\;\;\;\;\;\;\frac{1}{\pi}
\int_{-\pi}^{+\pi} \; d\phi \int_0^1 dr\, \,r\; |r\,\phi\rangle\langle r, \phi| = \I\;,\;\;
\]
where the ranges are
$(0 \le r \le 1,\,  -\pi \le \phi < \pi)$.
\vspace{.3cm}

~b) ~A Zernike basis in the space of square integrable functions  $L^2(\D)$ defined in 
$\D$.  
\vspace{.1cm}

Like in QM we find that the Hermite functions are a basis in $L^2(\R)$, starting from
the Zernike polynomials $\{R_n^m(r)\}$ we find that the Zernike functions 
$\{R_n^m(r)\, e^{{\bf i} m \phi}\}$  \cite{BoWo}.          
are a basis in the space of images  $L^2(\D)$.  

Following Dunkl \cite{Du} we change the Born and Wolf parametrization $\{n, m\}$ 
to $\{u,v\}$:
  \[
  u:=\frac{n+m}{2} \quad v:= \frac{n-m}{2}
  \]
that have ranges simpler then $\{n,m\}$, as they are independent natural numbers i.e. $\{u=0,1,2,\dots, \;\,v=0,1,2,\dots\}$.       
 We  introduce also a normalization factor to have the explicit orthonormalization we are 
  used in QM and we redefine the Zernike functions as:
 \[
Z_{u,v}(r,\phi) :=\; \sqrt{u+v+1}\;\, R_{u+v}^{u-v}(r)\; e^{{\bf i}(u-v)\phi}\,.
  \]
  
$\{Z_{u,v}(r,\phi)\}$  have the following properties:
\vspace{.1cm}

~1.   ~$Z_{u,v}(r,\phi)$ are square integrable 
in $L^2(\D, \,r \,dr \,d\phi) \equiv L^2(\D) $ . 

~2.    ~They  have the symmetries:
\[
Z_{v,u}(r,\phi) \,=\, Z_{u,v}^*(r,\phi) \,=\, Z_{u,v}(r,-\phi)\,.
\]

~3. ~They are orthonormal:
\[
\langle Z_{u',v'}| Z_{u,v}\rangle \;\,=\,\; \frac{1}{\pi}
\int_{-\pi}^{+\pi}  d\phi \int_0^1 dr\; r\; Z_{u,v}^*(r,\phi) \;Z_{u',v'} (r,\phi)
\,=\, \delta_{u,u'}\, \delta_{v,v'}\;\,.
\]

~4. ~They satisfy the completeness relation:
\[
\sum_{u,v=0}^\infty \,\, Z_{u,v}^*(r_1,\phi_1) \;Z_{u,v}(r_2,\phi_2) \;\,=\;\; \pi \;\,
\delta(r_1^{\;2}-r_2^{\;2})\;\, \delta(\phi_1-\phi_2)\;.
\]
Thus the Zernike functions $\{Z_{u,v}(r,\phi)\}$ are an orthonormal  basis for the square integrable functions
of the disk $L^2(\D)$ exactly as the Hermite
functions $\{\psi_n\}$ are an orthonormal basis for the square integrable functions 
defined in $\R$.
Any function\, $f(r,\phi) \in L^2(\D)$\, can be written, in the sense of convergence on 
the Hilbert space $L^2(\D)$, as superposition of $Z_{u,v}(r,\phi)$:
\[
f(r,\phi) \;=\; \sum_{u,v=0}^{\infty} f_{u,v}\; Z_{u,v} (r,\phi)
\] 
where $f_{u,v}$ are complex numbers given by
\[
f_{u,v} \;=\; \frac{1}{\pi}\; \int_{-\pi}^{+\pi} \; d\phi \int_0^1 dr \;r \; 
Z^*_{u,v} (r,\phi)\; f(r,\phi).
\]
Moreover 
\[
\langle f | f \rangle \;\;=\; \frac{1}{\pi}\; \int_{-\pi}^{+\pi} \; d\phi \int_0^1 dr \;r \;{|f(r,\phi)|}^2 \;= 
\sum_{u,v=0}^\infty |f_{u,v} |^2 \;< \infty\;.
\]
Note that we have assumed\, $\langle f| f\rangle < \infty$ but not\, $\langle f|f\rangle = 1$.
If we restrict $f(r,\phi)$ to be real (as it is usual in optics) we have\, $f_{v,u} = f_{u,v}^*$ .
\vspace{.3cm}

~c)  ~A discrete basis in the disk\, $\{|u,v\rangle\}$.  
\vspace{.1cm}

 We follow the same procedure of the line and introduce in the disk $\D$ 
 the discrete set $\{|u,v\rangle\}$ :
\[
|u,v\rangle \,:=\, \frac{1}{\pi}\; \int_{-\pi}^{+\pi} \; d\phi \; \int_0^1 dr\; r\; |r,\phi\rangle \;Z_{u,v}(r,\phi) \qquad 
(u,v = 0,1,2,\dots) .
\]
Like in QM, we have the orthonormal discrete relations:
\[
\langle u,v|u',v'\rangle = \delta_{uu'} \; \delta_{vv'} \,,\qquad \sum_{u,v}^\infty
|u,v\rangle \langle u,v| = { \I'} \;,
\]
where,  as $\{|u,v\rangle\}$ is a discrete basis and $\{|r,\phi\rangle\}$
a continuous basis, the two identities { $\I$} and { $\I'$} are identities in
different spaces as it happens on the line. 
Likewise\, $\{\psi_n\}$\, in QM, the Zernike functions are the transition matrices from
continuous and discrete basis:
\[
Z_{u,v}(r,\phi) \;=\; \langle r,\phi | u,v \rangle\,\,.
\]

~d) ~the disk $\D$ is an algebra representation of $su(1,1) \oplus su(1,1)$.  
\vspace{.1cm}

Analogously to the line, we introduce the discrete operators ${ U}$ and ${ V}$ 
such that
\[
U\, Z_{u,v}(r, \phi) = u\, Z_{u,v}(r, \phi)\,,\qquad  
V\, Z_{u,v}(r, \phi) = v\, Z_{u,v}(r, \phi)
\]
and the continuous ones $R$, $D_r$\, and\, $\Phi$\,  such that
\[
R\,Z_{u,v}(r, \phi)=r\, Z_{u,v}(r, \phi),\;\;\;\;  D_r\,Z_{u,v}(r, \phi)=\frac{\partial Z_{u,v}(r, \phi)}{\partial r},\;\;\;\; \Phi\,Z_{u,v}(r, \phi)=\phi \,Z_{u,v}(r, \phi)\,.
\]
Four additional operators on the\, $\{Z_{u,v}(r,\phi)\}$ define the recurrence relations:
%\[
%~~~~~~~~~{ {A}_+}\, Z_{u,v}(r, \phi)\;=\; ~~{ (u+1)}\, ~Z_{{ u+1},v}(r, \phi)\,,
%\]
%\[
%~~~~~~~~~ { {A}_-}\, Z_{{ u},v}(r, \phi)\;=\; ~~~~~~{ u}\, ~~~~~~Z_{{ u-1},v}(r, \phi)\,,
%\]
%\[
%~~~~~~~~{ {B}_+}\, Z_{u,{ v}}(r, \phi)\;=\; ~~{ (v+1)}\, ~Z_{u,{ v+1}}(r, \phi)\,,
%\]
%\[
%~~~~~~~~~~~{ {B}_-}\, Z_{u,{ v}}(r, \phi)\;=\; ~~~~~~{ v}\, ~~~~~~Z_{u,{ v-1}}(r, \phi)\; ;
%\]

\[
{ {A}_+}\, Z_{u,v}(r, \phi)\;=\; ~{(u+1)}\, ~Z_{{ u+1},v}(r, \phi)\,,
~~~~~~~~~ { {A}_-}\, Z_{u,v}(r, \phi)\;=\; ~{ u}\, ~Z_{{ u-1},v}(r, \phi)\,,
\]
\[
{ {B}_+}\, Z_{u,v}(r, \phi)\;=\; ~{(v+1)}\, ~Z_{u,{ v+1}}(r, \phi)\,,
~~~~~~~~~{ {B}_-}\, Z_{u,v}(r, \phi)\;=\; ~{ v}\, ~Z_{u,{ v-1}}(r, \phi)\; ;
\]
where

  \[
 ~~~A_+ ~:=~ \frac{e^{+{\bf i} \Phi}}{2} \sqrt{\frac{U+V+1}{U+V}}\; \left[-(1-R^2) D_R + R (U+V+2) +\frac{1}{R}(U-V)\right] \;,
  \]
  \[
  A_- ~:=~ \frac{ e^{-{\bf i} \Phi}}{2} \sqrt{\frac{U+V+1}{U+V}}\; \left[+(1-R^2) D_R + R (U+V) +\frac{1}{R}(U-V)\right]\;,~~~
  \]
  \[
 ~~~ B_+ ~:=~ \frac{ e^{-{\bf i} \Phi}}{2} \sqrt{\frac{U+V+1}{U+V}}\; \left[-(1-R^2) D_R + R (U+V+2) -\frac{1}{R}(U-V)\right]\;,
  \]
  \[
  B_- ~:=~ \frac{ e^{+{\bf i} \Phi}}{2} \sqrt{\frac{U+V+1}{U+V}}\; \left[+(1-R^2) D_R + R (U+V) -\frac{1}{R}(U-V)\right]\;. ~~~ 
  \]

The relevant commutation relations are :
\[
[U,V] = 0, \quad[U, A_\pm] = \pm A_\pm, \quad [U, B_\pm] = 0,
\]
\[
 [V, A_\pm] = 0, \quad [ V, B_\pm ] = \pm B_\pm ,\quad [A_\pm, B_\pm]=0 ,\quad [A_\pm, B_\mp]=0 \,;
\]
and, on the $\{Z_{u,v}(r,\phi)\}$ :
  
\[
[A_+,A_-]\, Z_{u,v}(r,\phi)\; =\;  -2(u+1/2)\; Z_{u,v}(r,\phi) \,;
\]
defining therefore {$A_3 := U+1/2$} we find that\, { $\{A_+,\, A_3,\, A_-\}$}\, are the generators of one algebra $su(1,1)$, we call { $su_A(1,1)$}:
\[
{ [A_+,\, A_-]\;=\;-2 A_3 \qquad [A_3,\, A_\pm]\,=\,\pm A_\pm}\, .
\]
Analogously
\[
[B_+,B_-]\; Z_{u,v}(r,\phi)= -2(v+1/2)\; Z_{u,v}(r,\phi) \,
\]
exhibits that  ${ \{B_+, \,B_3:= V+1/2\; ,\, B_-\}}$ are the generators of one other
algebra $su(1,1)$, we call { $su_B(1,1)$}\,:
\[
{ [B_+,\, B_-]\,=\,-2 B_3 \qquad [B_3,\, B_\pm]\,=\,\pm B_\pm}\, .
\]
Finally, as 
\[
{ [A_i, B_j] = 0\, ,}
\]
we obtain on the disk\, $\D$\, a differential representation
of the 6 dimensional Lie algebra\, { $su_A(1,1)\oplus su_B(1,1)$}\,.

We can now calculate the values of the Casimir invariants:
\[
\C_A\;  Z_{u,v}(r,\phi)\,=\,\left[A_3^2 -\frac{1}{2} \{ A_+, A_- \}\right]\, Z_{u,v}(r,\phi)\, =\, -\frac{1}{4}\, Z_{u,v}(r,\phi)\,,
\]
\[
\C_B\;  Z_{u,v}(r,\phi)\,=\,\left[B_3^2 -\frac{1}{2} \{ B_+, B_- \}\right]\, Z_{u,v}(r,\phi)\, =\, -\frac{1}{4}\, Z_{u,v}(r,\phi)\,.
\]
As the Casimir of the discrete principal series of\, $su(1,1)$ is $j(j+1)$ with 
$j=-1/2, -3/2, \dots$, the\, RHS\, $\{Z_{u,v}(r,\phi)\}$\, is isomorphic to the fundamental
{ unitary irreducible representation} { $D^+_{-1/2} \otimes D^+_{-1/2}$} 
of { $SU(1,1)\otimes SU(1,1)$}
where the eigenvalues of $U$ and $V$ are the natural numbers and the
eigenvalues of $A_3$ and $B_3$ are the half-integers numbers $1/2,\, 3/2,\, 5/2,
\dots $ \cite{Bar}.
\vspace{.2cm}

%%%%%%%%%%%%%%%%%%%%%%%%%%%%%
\section{ ~Operators, UEA and Lie group in the space of Images} 
%%%%%%%%%%%%%%%%%%%%%%%%%%%%%
\vspace{.1cm}

The Universal Enveloping Algebra of\, $su(1,1)\oplus su(1,1)$
{ UEA[$su(1,1)\oplus su(1,1)$]} is the algebra constructed on the ordered
monomials $A_+^{\,\alpha} A_3^{\,\beta} A_-^{\,\gamma}
B_+^{\,\delta} B_3^{\,\epsilon} B_-^{\,\zeta}$ where\, $\alpha, \beta,
\gamma, \delta, \epsilon$ and $\zeta$\, are natural numbers.
Analogously to\, $\{\psi_n\}$:

  - $ \{Z_{u,v}(r,\phi)\}$ is an irreducible representation so that every linear operator
  ${\cal O}$ acting on $\D$ and $L^2(\D)$  belongs to UEA[$su(1,1)\oplus su(1,1)$].
   
  - $\{Z_{u,v}(r,\phi)\}$ is a unitary representation so that every base 
  in the space\, $L^2(\D)$\, has the form  ~$\{g\, W_{u,v}(r,\phi)\}$
where $g$ is an arbitrary element of the group { $SU(1,1) \otimes SU(1,1)$}\,.
\vspace{.2cm}

%%%%%%%%%%%%%%%%%%%%%%%%%
\section{~~Measure in the space of Images}
%%%%%%%%%%%%%%%%%%%%%%%%%
\vspace{.1cm}

Measure does not belong to the theory of the spaces. It is something added
in applications to the mathematical formalism.
HS and RHS are completely independent
from the measure and different measures describe different physics.
We have discussed the probabilistic QM measure. 
As Theory of Images is a classical theory its measure is deterministic:
mathematically is a functional\, i.e. a mapping ${\cal M}$  from the images into the real numbers:
\[
{\cal M}\left[|f\rangle \right]  = \; m\left[|f\rangle \right] \,\in \,{\bf R}\,.
\]

For example, the luminosity ${\cal L}$ of the image $|f\rangle$ could be defined as the number
\[
 {\cal L}\left[|f\rangle\right]\; :=\;\frac{1}{\pi}\;  \int_{-\pi}^{+\pi} d\phi  \int_0^1 dr\; r\;\,  {\left|f(r,\phi)\right|}^2\, .
 \]
 
%%%%%%%%%%%%%%%%%%%
\section{Conclusions}
%%%%%%%%%%%%%%%%%%%
\vspace{.1cm}

~~~~~- ~RHS describe both quantum and classical theories.

~-~ In RHS we have continuous and discrete variables that, through the 
related lowering and rising operators, define a Lie algebra.

~- ~The space of the wave functions is an unitary irreducible representation of this algebra.

~- ~Elements of the group define the bases of this space.

~- ~UEA is a basis in the space of operators acting on the system.

~- ~Classical measures are described by functionals. 
\vspace{.2cm}

%%%%%%%%%%%%%%%%%
%\section*{References}
%%%%%%%%%%%%%%%%%
%\vspace{.3cm}

\end{document}